\documentclass[twocolumn,preprintnumbers,amsmath,amssymb]{revtex4}

\def\ga{\mathrel{\raise.3ex\hbox{$>$\kern-.75em\lower1ex\hbox{$\sim$}}}}
\def\la{\mathrel{\raise.3ex\hbox{$<$\kern-.75em\lower1ex\hbox{$\sim$}}}}

\newcommand\beq{\begin{equation}}
\newcommand\eeq{\end{equation}}
\newcommand\beqar{\begin{eqnarray}}
\newcommand\eeqar{\end{eqnarray}}
\usepackage{graphicx}
\usepackage{epsfig}
\usepackage{epsf}
\usepackage{rotating}
\usepackage{verbatim}

\begin{document}

\preprint{ArXiv:/0809.2779}
\preprint{UMN--TH--2717/08}
\vskip 0.2in
\title{Instability of anisotropic cosmological solutions supported by vector fields.}

\author{Burak Himmetoglu$^{(1)}$, Carlo R. Contaldi$^{(2)}$ and Marco Peloso$^{(1)}$}
\address{$^{(1)}${\it School of Physics and Astronomy, University of Minnesota, Minneapolis, MN 55455, USA}}
\address{$^{(2)}${\it Theoretical Physics, Blackett Laboratory, Imperial College, London, SW7 2BZ, UK}}

\begin{abstract}
Models with vector fields acquiring a non-vanishing vacuum expectation value along one spatial direction have been proposed to sustain a prolonged stage of anisotropic accelerated expansion. 
Such models have been used for realizations of early time inflation, with a possible relation to the large scale CMB anomalies, or of the late time dark energy. We show that, quite generally, the concrete realizations proposed so far are plagued by instabilities (either ghosts, or unstable growth of the linearized perturbations) which can be ultimately related to the longitudinal vector polarization present in them. Phenomenological results based on these models are therefore unreliable.
\end{abstract}

\maketitle

{\it Introduction.}~~Observations of the Cosmic Microwave Background
(CMB) anisotropies in the WMAP experiment~\cite{WMAP} are in overall
agreement with the inflationary paradigm. However, certain features of
the full sky maps seem to be anomalous in the standard
picture. These anomalies include the low power in the quadrupole
moment~\cite{cobe,wmap1,lowl}, the alignment of the lowest multipoles,
also known as the `axis-of-evil'~\cite{axis}, and an asymmetry in
power between the northern and southern ecliptic
hemispheres~\cite{asym}. The statistical significance of these effects has been 
debated in the literature.  The discussion is complicated by the
difficulty of quantifying the {\sl a posteriori} probability of the
effects in the different maps produced by foreground cleaning methods
and in the context of statistical anisotropy where the use of the
angular power spectrum as a statistic can be misleading. However,
recent studies on properly masked data have shown that an anisotropic
covariance matrix fits the WMAP low-$\ell$ data at the 3.8$\sigma$
level~\cite{Groeneboom:2008fz} \footnote{The anisotropy found there is at the $10\%$ level.
The study of the quadrupole moments of the power spectrum rules out an ${\rm O } \left( 1 \right)$ anisotropy \cite{ArmendarizPicon:2008yr}.}. The significance of the anomalous lack
of large-angle correlations, together with the alignment of power has
also grown in strength with the latest data~\cite{Copi:2008hw}, with
only one in 4000 realizations of the concordance model in agreement
with the observations.

Such violations of statistical isotropy are considered at odds with
the standard phase of early inflation. However, an
albeit more plausible explanation of these anomalies arising from a
systematic effect or foreground signal affecting the analysis is not
forthcoming.  This has led to a number of attempts at reconciling some
of the anomalies with the standard inflationary picture through
various modifications.  Specifically, the alignment of lowest
multipoles could be related to an {\sl anisotropic} inflationary era,
whose duration is fine tuned so that the signature will be observed in
the modes entering the horizon today, thus modifying the lowest
multipoles~\cite{gcp1,gcp2,uzan1,uzan2}. An anisotropic expansion has
also been considered for late time acceleration~\cite{mota}. Although
the present statistics of the observed Supernovae does not show any
evidence for the anisotropy~\cite{SN}, such studies are motivated by
the large increase of data that is expected in the next few years,
with surveys returning many thousands of SN1a light curves over
thousands of square degrees, and by the fact that one should keep an
open mind on the nature of dark energy given our present lack of
understanding.

Anisotropic, but spatially homogeneous backgrounds were classified
into equivalence classes long ago by Bianchi~\cite{bianchi}. In the
presence of a cosmological constant and matter fields satisfying strong
and dominant energy conditions, all Bianchi models, with the 
possible exception of Bianchi-IX, undergo a rapid
isotropization~\cite{wald}.  The full formalism for cosmological 
perturbations in Bianchi-I backgrounds (the simplest of these classes)
has been recently developed in~\cite{gcp1,gcp2,uzan1,uzan2}, with an
application to the case in which the only source is a slowly rolling inflaton field,
which causes the isotropization as an effective cosmological constant. This is the
simplest anisotropic scenario which can be associated with a later inflationary stage.
In principle, such a background solution can have striking signatures, provided that
the following inflationary stage is not too long. 
Specifically, different $a_{\ell m}$ coefficients of the CMB multipole 
expansion are correlated to each other, and the two gravity wave (GW) polarizations
behave in a nonstandard manner and can differ from each other \cite{gcp1,gcp2}.
In the case of axisymmetric expansion (equal expansion rate in two directions),
one of the two GW polarizations experiences a large growth during the anisotropic era, 
which may result in a large $B$ signal in the CMB \cite{gkp}. An analogous growth is
expected also for the general (non axisymmetric) case.

Such simple models, however, do not allow for a small and controllable departure from
isotropy. Indeed, the isotropization due to the (effective) cosmological
constant starts from a (Kasner-type) singularity \cite{gkp} and lasts only for about 
one e-fold ($\Delta t \sim H^{-1} \,$, where $H$ is the expansion rate due to the cosmological constant). 
As a consequence, one loses predictive power on the initial conditions for the system.
A prolonged anisotropic stage can be obtained by introducing some ingredients that violate the
premises of Wald's theorem~\cite{wald} on the rapid isotropization of
Bianchi universes.  This has been realized through the addition of
quadratic curvature invariants to the gravity action~\cite{barrow},
with the use of the Kalb-Ramond axion~\cite{nemanja}, or of vector
fields~\cite{ford}. In this Letter, we focus on this last possibility,
as it is perhaps the simplest one (at least, from a technical point of
view).  The evidence for an anisotropic covariance matrix reported
in~\cite{Groeneboom:2008fz} is based on a primordial power spectrum
for the perturbations which is motivated by one of such models
\cite{acw}; therefore, such constructions deserve close scrutiny.

In these models, a vector field with non-vanishing spatial vev is
responsible for the anisotropy. To our knowledge, there are three
different realizations of this mechanism. The oldest one dates back to
1989 \cite{ford}, and the vev of the vector field is due to a
potential $V \left( A_\mu \, A^\mu \right)$ involving only the vector
$A_\mu$. A more recent proposal is characterized by a non-minimal
coupling $R \, A_\mu \, A^\mu$ of the vector field to the
curvature \footnote{Such a coupling was suggested in ref. \cite{Turner:1987bw} as
a mechanism to generate primordial magnetic fields.}. 
For a special value of this coupling, the vev of $A_\mu$ can have a slow roll
evolution. While the original proposal of this idea \cite{mukvect} realizes an
inflationary background through several vector fields,
ref. \cite{soda1} suggested a simplified version in which an inflaton
scalar field is the main source of expansion, while the vector field
supports the anisotropy. A completely different class of models makes
use of a Lagrange multiplier to force a space-like fixed vev for the
vector field~\cite{acw}~\footnote{Models with vector fields
spontaneously breaking Lorentz invariance were introduced in
\cite{kosa}.  For review and references on models with
time-like fixed vev vectors see \cite{jacrev}.}.  These three
different implementations have been realized and studied by several
authors \cite{mota,vectvev,dgw,deltaN}.

We show that these three class of models contain instabilities which
did not emerge in previous studies. We see this from the linearized
study of the perturbations 
around the anisotropic inflationary solutions of these models. As in
all slow roll inflationary backgrounds, each mode of the perturbations
is initially in the small wavelength regime (the wavelength is
exponentially small at early times); as the background inflates, the
wavelength becomes larger than the Hubble horizon $H^{-1}$
\footnote{For anisotropic backgrounds, there are different expansion
  rates $H_i$ for the different directions; however, in the
  phenomenologically relevant cases of small anisotropies the
  different expansion rates parametrically coincide.} and the mode
enters the large wavelength regime. This transition is dubbed
horizon crossing. For the model of \cite{ford}, the system of
perturbations contains a ghost in the small wavelength regime. For the
case of the non-minimal coupling with curvature of \cite{mukvect,soda1},
and the fixed-norm case of \cite{acw}, the ghost appears from some
interval of time close to horizon crossing \footnote{Such instabilities
do not show up in studies based on the $\delta N$ formalism
\cite{deltaN}, since such formalism gives the evolution of the perturbations
only after horizon crossing.}. The system of linearized
perturbations blows up at this moment\footnote{Ref. \cite{Clayton:2001vy}
also showed that systems of the type \cite{acw} have unstable
solutions in Minkowski spacetime.}.

Complete computations of cosmological perturbations are rather
tedious, and the results for the present cases can be obtained only
through involved algebra. We have performed these
computations along the lines of \cite{gcp1,gcp2}. We first write the
most general system of perturbations (both of the metric and of the vector field); 
we then fix the freedom of
general coordinate invariance, we integrate out the non-dynamical
modes, and finally we study the remaining system of dynamical
perturbations. The divergence of the linearized perturbations is found
by solving the linearized Einstein equations. 
Ghost are found from studying the kinetic matrix that
couples the dynamical perturbations in their quadratic action. Such
computations cannot be reported in this Letter, and due to their length,
do not provide an insight on the true nature of the problem. For
this reason, we report them in a separate and more extended publication
\cite{hcp2}. The fact that the instability is related to the vector
field, fortunately suggests that a partial study, with only the
perturbations of this field included, can shed light on
the true nature of the problem, without the need to go through the technicalities
of the full computation. The results of this analysis, which are 
summarized in the next Section, show that this is indeed the case.
The significance of these results is discussed in the concluding Section.

 \smallskip {\it The instabilities.}~~ We assume that the spatial vev
 of the vector field is aligned along the $x$ direction, $\langle A_x
 \rangle \neq 0 \,$, so that the line element is
\begin{equation}
d s^2 = - d t^2 + a \left( t \right)^2 d x^2 + b \left( t \right)^2 \left[ d y^2 + d z^2 \right]\,.
\label{line}
\end{equation}
We introduce the two expansion rates $H_a \equiv \dot{a} / a ,\, H_b
\equiv \dot{b} / b \,$, and we define their average $H$ and rescaled
difference $h$ through $H \equiv \frac{H_a + 2 \, H_b}{3}$ and $ h \equiv  \frac{H_b-H_a}{3}\,$.
The inflationary expansions that we consider below are characterized
by constant or slowly evolving rates. For the models we are
considering, $h / H = {\rm O } \left( B^2 \right) \,$, where $B$ is
the rescaled vev of the vector field $\langle A_x \rangle \equiv M_p \, a \, B \,$
\cite{ford,soda1,acw}. Therefore, $B$ must also be slowly rolling
during the slow roll regime. We consider the phenomenologically relevant case
of moderate anisotropy, $B < 1 \,$.

Before studying these models, consider a massive vector field in an isotropic background (eq.~(\ref{line}) with $a=b$)
\begin{equation}
S = \int d^4 x \, \sqrt{-g} \, \left[ - \frac{1}{4} \, F_{\mu \nu} \, F^{\mu \nu} - \frac{M^2}{2} \, A_\mu \, A^\mu \right]\,.
\label{actmass}
\end{equation}
We assume that $A_\mu$ has vanishing vev, and we decompose its fluctuations as 
$A_\mu = \left( \alpha_0 ,\, \partial_i \alpha_L + \alpha_i^T \right) \,$.
The transverse vector perturbation $\alpha_i^T$, satisfying $\partial_i \alpha_i^T = 0 \,$, contains two physical modes. These modes are well behaved, and decoupled from the $\alpha_0 ,\, \alpha_L$ perturbations. We disregard them in the following. For $M^2 \neq 0 \,$, the two perturbations $\alpha_0 ,\, \alpha_L$ encode one additional degree of freedom, namely the longitudinal vector polarization. Indeed the mode $\alpha_0$ is non-dynamical, since it appears without time derivatives in the action, and must be integrated out. Namely, its equation of motion, after Fourier decomposition in the spatial directions, gives $\alpha_0 = \left[ p^2 / \left( p^2 + M^2 \right) \right] \dot{\alpha}_L \,$, where $p = k / a$ is the physical momentum of the mode, $k$ the comoving momentum, and dot denotes time differentiation. Inserting this solution back into (\ref{actmass}) we obtain the action for the dynamical mode. In Fourier space it reads
\begin{equation}
S_{\rm longitudinal} = \int  d t \, d^3 k \, a^3 \: \frac{p^2 \, M^2}{2} \left[ \frac{\vert \dot{\alpha_L} \vert^2}{p^2 + M^2} - \vert \alpha_L \vert^2 \right]\,.
\end{equation}
The longitudinal vector mode exists due to the mass term, so it is not a surprise that $M^2$ multiplies the kinetic term. We see that this mode is a ghost for $M^2 < 0 \,$.

Let us now turn to the models of our interest. The two models \cite{ford} and \cite{soda1} can be studied together, using \footnote{The cosmological expansion in \cite{ford} is also driven by a cosmological constant, which, in our notation, is included in the potential $V \left( A^2 \right) \,$. Ref. \cite{soda1} considered only a quadratic term in $V \left( A^2 \right)$ (therefore, only $\frac{\partial V}{\partial A^2}$ is non-vanishing; this does not affect our analysis)
and also introduced a slowly rolling inflaton field. For the present study, the inflaton can be replaced by a cosmological constant; the exact model of \cite{soda1} is studied in \cite{hcp2}.}
\begin{equation}
S = \int d^4 x \, \sqrt{-g} \, \left[ \frac{M_p^2}{2} \, R - \frac{F^2}{4} - V \left( A^2 \right) + \frac{\xi}{2} \, R \, A^2 \right]\,.
\label{actf-s}
\end{equation}
Expanding the potential at quadratic order in $A_\mu$, and comparing with eq.~(\ref{actmass}), this action leads to the mass term
\begin{equation}
M^2 = 2 \, \frac{\partial V}{\partial A^2} - \xi \, R = 2 \, \frac{\partial V}{\partial A^2} - 6 \xi \left(
2 H^2 + h^2 + \dot{H} \right)\,.
\label{mass}
\end{equation}
The equations of motion for the rescaled vev $B$ obtained from (\ref{actf-s}) is
\begin{eqnarray}
&&\!\!\!\!\!\!\!\!\!\!\!\!\ddot{B} + 3 \, H \, \dot{B} + {\cal Q} B = 0\,, \\
&&\!\!\!\!\!\!\!\!\!\!\!\!{\cal Q} \equiv 2 \, \frac{\partial V}{\partial A^2} - 2 \, H \, h - 5 \, h^2 - 2 \, \dot{h} 
+ \left( 1 - 6 \, \xi \right) \left( 2 \, H^2 + h^2 + \dot{H} \right)\,. \nonumber
\end{eqnarray}

Slow roll of $B$ requires ${\cal Q} \ll H^2 \,$ (since the $3 H
\dot{B}$ term provides a ``friction'' to the motion). This is achieved
in two different ways by \cite{ford}  and \cite{soda1}. Ref.~\cite{ford} studied solutions with constant $H_{a,b}$ in absence of the $A^2 \, R$ term, $\xi = 0 \,$. This requires ${\cal Q} = 0$, or, in other terms
\begin{equation}
\frac{\partial V}{\partial A^2} = - H^2 + H \, h + 2 \, h^2 = - H_a \, H_b < 0\,.
\end{equation}
This corresponds to a negative square mass in eq.~(\ref{mass}). From
our discussion of the model~(\ref{actmass}) we therefore immediately
see that the longitudinal vector polarization is a ghost in the limit
of isotropic background ($B = 0$). In \cite{soda1}, the choice $\xi =
1/6$ is made, so that (following the idea of \cite{mukvect}) the ${\rm
  O } \left( H^2 \right)$ contribution is absent from ${\cal Q}
\,$. Then, slow roll is achieved in the case of small anisotropy, $B \ll
1 \,$, and for $\partial V / \partial A^2 \ll H^2 \,$. We then see
that the square mass parameter (\ref{mass}) is negative in this limit, 
indicating that the longitudinal vector polarization is a ghost
in the isotropic limit in this case too. A more detailed study, 
including also metric perturbations, shows that the ghost persists also for
moderate anisotropy \cite{hcp2}.

Finally, let us discuss the stability of model \cite{acw},
\begin{equation}
S = \int d^4 x \, \sqrt{-g} \, \left[ \frac{M_p^2}{2} \, R - \frac{F^2}{4} + \lambda \left( A^2 - m^2 \right) - V_0 \right]\,.
\label{actf-acw}
\end{equation}
In this case, the rescaled vev $B$ is forced to be constant, and equal
to $m / M_p$ by the Lagrange multiplier $\lambda \,$. One then finds a
background solution with constant expansion rates satisfying $H_b
= \left( 1 + B^2 \right) H_a \,$ \cite{acw}. We decompose the vector
field in vev plus fluctuations, $A_\mu = \langle A_\mu \rangle +
\left( \alpha_0 ,\, \alpha_1 ,\, \partial_i \alpha + \alpha_i \right)$
where the index $i = 2 ,\, 3$ spans only the coordinates of the $y-z$
plane, and $\partial_i \alpha_i = 0 \,$. The equation of motion for
$\lambda$, once expanded at the linearized level in the
perturbations, reads $B \, \alpha_1 = 0 \,$. This equation identically
vanishes if the background is isotropic ($B=0$), while it eliminates
one of the vector perturbations for $B \neq 0 \,$. Therefore, contrary
to the previous study, we cannot consider the isotropic limit in this
case. 

We are interested in the quadratic action for the perturbations.
It is easy to see that the perturbations $\alpha_i$
decouple. We are then left with a quadratic action containing
$\alpha_0$ and $\alpha$. $\alpha_0$ is non-dynamical in this case too,
and can be integrated out, leading to the quadratic action
\begin{eqnarray}
\delta^2 S \!\!\!&=&\!\!\! \frac{1}{2} \int d t \, d^3 k \, a \, b^2 p_T^2 \left( p_L^2 - 2 \, H_a \, H_b \right) \nonumber\\
&& \quad\quad \times \left[ \frac{\vert \dot{\alpha} \vert^2}{p_L^2 + p_T^2 - 2 \, H_a \, H_b} - \vert \alpha \vert^2 \right]\,,
\end{eqnarray}
where $p_L$ and $p_T$ are the components of the physical momentum
along the $x-$direction and in the perpendicular $y-z$ plane,
respectively. We see the presence of a ghost close to horizon crossing
(we recall that $H_a$ and $H_b$ are constant, while the physical
momentum exponentially decreases, $p_L \propto a^{-1} ,\, p_T \propto
b^{-1} \,$). Moreover, the equation of motion for $\alpha$ (and the
corresponding solution) diverges when the prefactor $p_L^2 - 2 \, H_a
\, H_b$ vanishes \footnote{Clearly, this equation in one of the linearized equations for the
perturbations, and can be obtained without the need of computing the quadratic action.
Once also metric perturbations are included, this instability shows up from the linearized
Einstein equations \cite{hcp2}. An analogous instability arises in \cite{soda1} for moderate anisotropy.}. These instabilities are confirmed by the full
computation of \cite{hcp2}.

\smallskip {\it Discussion.}~~ We start by stressing the limits of our computation. The above
results have been obtained for standard kinetic terms for the vector
field. Since the U(1) symmetry is
anyhow broken by the potential term, 
there is no special reason for this choice. Indeed, works on lorentz
violating vector fields study generalized kinetic terms of the type
${\cal L} \supset - \beta_1 \, \nabla^\mu A^\nu \nabla_\mu A_\nu -
\beta_2 \left( \nabla_\mu A^\mu \right)^2 - \beta_3 \nabla^\mu A^\nu
\nabla_\nu A_\mu \,$. The standard kinetic term corresponds to 
$\beta_1 = - \beta_3 = 1/2 ,\, \beta_2 = 0
\,$. Ref. \cite{ford} only discusses the case of a standard kinetic
term. We have studied perturbations in this model for arbitrary
$\beta_i$ coefficients. We find that the ghost is absent for $\beta_1
+ \beta_2 + \beta_3 \neq 0 \,$. However, in this case one of the
perturbations is a tachyon in the early time/small wavelength
regime \cite{hcp2}. Ref. \cite{dgw} showed that the model \cite{acw} is unstable
in the case of $\beta_1 + \beta_2 + \beta_3 \neq 0
\,$ \footnote{Ref. \cite{dgw} also studied the case $\beta_1 +
\beta_2 + \beta_3 =0$, which includes the standard kinetic
term. However, computations were performed only in the
small or large wavelength limit, and therefore could not find the
instability appearing at horizon crossing.}. For the non-minimal coupling to
the curvature, all the studies done so far are limited to standard
kinetic terms, and therefore our computations have also been
restricted to this case. A second limitation is related to the fact
that we have performed specific computations only for the three models
\cite{ford,acw,soda1}. However, these models 
are ``prototypes'' for the three different ways of obtaining the non-vanishing spatial
vev explicitly realized in the literature. We expect that the issues raised here also arise 
in all models for which the anisotropy is obtained in one of these three ways.

Due to these limitations, we do not claim that all possible models of
anisotropic expansion through vector fields are ruled out by our
findings. Nonetheless, the issues we have found are specific to models
with vector fields, and should be checked in the stability analysis of
all these models. Ghosts or tachyons in the early time/small wavelength
regime are an indication that the vacuum of the model is unstable;
if these instabilities were present  only at an energy scale $\Lambda$ (inverse
wavelength) much greater than the hubble scale $H$, we could simply
consider these theories as effective ones, valid below that scale $\Lambda
\,$. Cosmological perturbations could therefore be treated as in the
standard case, and they would lead to predictive results that can be
trusted up to ${\rm O } \left( H / \Lambda \right)$ corrections
(possibly, to some higher power). The fact that the instabilities we
have found persist up to horizon crossing indicate that $\Lambda$ in
these cases should be comparable or smaller than the Hubble
scale. Therefore, even assuming that these models can be ``cured'',
the resulting phenomenology would be profoundly different. Although
our concrete studies have produced a negative outcome, we hope that
some specific constructions will eventually be able to overcome the
issues found here.

\smallskip {\it Acknowledgments.}~~We thank , L. Kofman, D.F. Mota, E. Poppitz, J. Soda, M.B. Voloshin, and particularly N. Kaloper for very useful discussions. The work of
B.H. and M.P. was partially supported by the DOE grant
DE-FG02-94ER-40823.

\end{document}